# High sensitivity refractive index sensor based on the semicircular bent fiber

Tingting Wu,[1] Linlin Xu,[1] and Xinhai Zhang [1,a]

[1] *Department of Electrical and Electronic Engineering, Southern University of Science and Technology, Shenzhen 518055, China*

A refractive index sensor based on a semicircular bent fiber is presented. The interference occurs between the cladding mode excited in the bending region and the core mode. Both the theoretical and experimental results show that the resonant dip wavelength decreases linearly with the increase of the refractive index of the surrounding environment. A high sensitivity of 1031 nm per refractive index unit is obtained over the refractive index range of 1.3324 to 1.3435 by using a bent fiber with a bending radius of 500 μm.

Recently, sensors based on the bent fiber have been extensively investigated due to their easy fabrication method and simple configuration. Acoustic, temperature, displacement sensors have already been reported using bending-induced birefringence in a fiber.[1-6] Evanescent wave sensor[7] and refractive index (RI) sensor[8] have also been realized based on the bending-induced transmission loss. However, loss modulated sensors are easily affected by light source or external environment, which limited their applications. To solve the problem and take the advantage of the cladding mode of the fiber is sensitive to the RI of the surrounding environment, interference effect between the cladding mode and the core mode or different cladding in a bent fiber has been exploited for RI sensing applications.[9-14] For examples, Luo et al. proposed an RI sensor based on a C-shaped ultrathin fiber taper,[10] a sensitivity of 658 nm/RIU (refractive index unit) for an RI range of 1.333 to 1.353 was achieved. Liu et al. proposed an 'S'-like tapered fiber and the sensitivity as high as 4000 nm/RIU was obtained over an RI range of 1.424 to 1.435 [11]. To eliminate the temperature cross-sensitivity, Zhang et al. demonstrated an RI sensor using a hybrid structure with an LPG and a bent-fiber intermodal interferometer. The sensitivity of the structure is 183.44 nm/RIU at an RI range of 1.3269 to 1.3721.[12] Even though the sensors based on the bent tapers have higher sensitivities, thinner waist diameter of the tapers would cause the structure more fragile. In this letter, an RI sensor based on the semicircular bent structure was proposed, the sensitivity of the structure increases with the increasing of the bending radius.

Figure 1 shows the schematic of the light transmitted in the bent fiber structure. Before entering into the bending region, light guided in the fiber is mainly confined in the fiber core, which is usually called core mode. After entering into the bending structure, part of light in the core mode will be coupled into the fiber cladding mode because of the decreasing of the incident angle of the core mode at the interface of the fiber core and fiber cladding. And at the interface of the fiber cladding and the surrounding environment, part of the cladding mode will be leaked into the surrounding environment, which mainly contributes to the transmission loss of the bending fiber structure. The cladding mode will be coupled back to the fiber core when leaving the bending structure. Because the effective refractive indices (ERIs) of the core mode and cladding mode are different, interference will be formed after the cladding mode is coupled back to the core of the fiber.

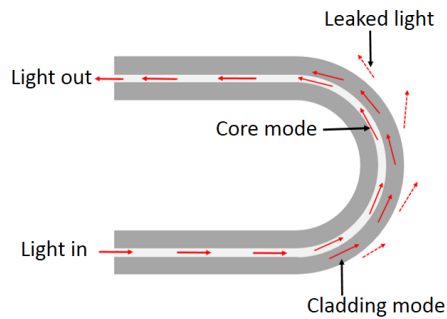

FIG. 1. Schematic of the light transmitted in the sharply bent fiber.

a) Electronic mail: zhangxh@sustc.edu.cn.

The phase difference φ between the core mode and cladding mode can be expressed as

$$\varphi = k L (n_{co}^{eff} - n_{cl}^{eff}), \qquad (1)$$

where $k$ is the wave number, $L$ is the bending length of the structure, $n_{co}^{eff}$ and $n_{cl}^{eff}$ are the ERIs of the core mode and cladding mode, respectively. The cross section of the fiber is shown in Figure 2. The resonant dip wavelength $\lambda_{dip}$, where light intensity is lowest, can be expressed as

$$\lambda_{dip} = 2 L (n_{co}^{eff} - n_{cl}^{eff})/(2m + 1), \qquad (2)$$

where the $m$ is an integer.

Since $n_{co}^{eff}$ is almost unaffected by the RI $n_{en}$ of the surrounding environment, $n_{co}^{eff}$ can be assumed to be a constant value when $n_{en}$ changes. However, $n_{cl}^{eff}$ is sensitive to the changing of $n_{en}$, which leads to the shift of $\lambda_{dip}$ when $n_{en}$ changes.

The theoretical sensitivity $S_{th}$ of the bent structure is then can be calculated according to

$$S_{th} = d\lambda_{dip}/dn_{en} = -2L dn_{cl}^{eff}/dn_{en}/(2m + 1), \qquad (3)$$

where, $dn_{co}^{eff}/dn_{en}$ is neglected since $n_{co}^{eff}$ almost keeps constant when $n_{en}$ changes. It can be seen from Eq. (3) that, $\lambda_{dip}$ shifts to a shorter wavelength with the increasing of $n_{en}$. According to Eq. (2) and Eq. (3), both the sensitivity of the sensor and the resonant wavelength increase linearly with the increasing of bent length.

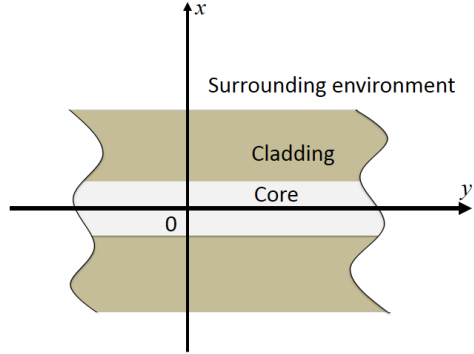

FIG. 2. The cross section of the fiber.

To verify the theoretical predictions, in experiment, the semicircular bent structures with different bending radius $R$ were fabricated, and their responses to the changing of RI $n_{en}$ of the surrounding environment were also tested. The schematic of the fabrication of the sharply bent fiber structure is shown in Figure 3(a). The bent structure was fabricated on a standard telecommunication single-mode fiber (Corning SMF-28e), as mentioned above, by flame-heated treatment. It should be emphasized that before the heating, the jacket surrounding the bare fiber with a length of 5 mm should be totally stripped. During the heating process, one end of the bare fiber was bent around a solid cylinder with a diameter of several hundred micrometers at an angle of 180° while the other end was fixed on a stage. At the same time, the transmission spectrum of the SCF was monitored by an optical spectrum analyzer (OSA) with a broadband light source (BBS) with a wavelength window from 1500 to 1600 nm used as the input light source. The optical microscopic image of the semicircular bent fiber structure is shown in Figure 3(b).

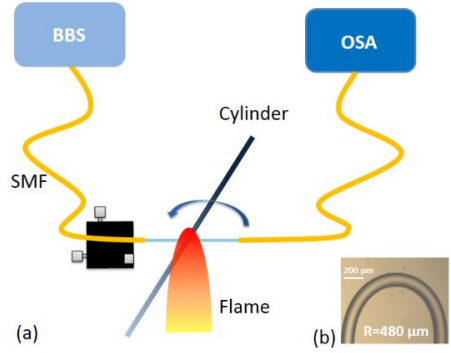

FIG. 3. (a) Schematic of the fabrication of the sharply bent fiber structure. (b) Optical microscopic image of the fabricated structure with a bending radius and angle of 402 μm and 180°, respectively.

Glycerol aqueous solutions with different RIs were used to test the RI $n_{en}$ response of the fabricated structure. Six volume ratios of pure water and pure glycerol, 10:0, 99:1, 98:2, 97:3, 96:4, and 95:5, were chosen in preparing the solutions. The volume of water and glycerol were well controlled by using a syringe. The RIs $n_{en}$ of the solutions can be calculated according to Gladstone-Dale relation[15]

$$n - 1 = \Phi_1(n_1 - 1) + \Phi_2(n_2 - 1), \quad (4)$$

where $\Phi_1$ and $\Phi_2$ are the volume fractions of the glycerol and water, $n_1 = 1.3324$ and $n_2 = 1.4715$ are refractive indices of glycerol and water at 589 nm at room temperature of 25 ℃. The calculated RIs are 1.3324, 1.3338, 1.3352, 1.3366, 1.3380, and 1.3394, respectively. Figure 4 shows the transmission spectra of the bend structure with $R = 480$ μm that immersed into the six different solutions. The recorded transmission spectra were obtained by subtracting the spectrum of the light source from the transmission spectra of the bent fiber. It should be noted that before each measurement, the structure was cleaned by using alcohol to remove the residual glycerol solution.

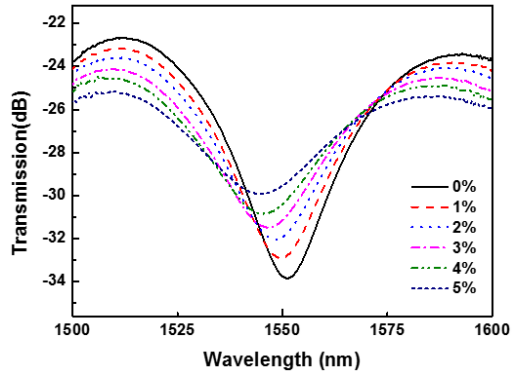

FIG. 4. Transmission spectra of the sharply bent fiber with R = 480 μm that immersed into glycerol solutions with different volume rations.

As shown in Figure 4, when the structure is immersed in pure water, $\lambda_{dip}$ is observed to be 1551.2 nm for the light window used in the experiment. The depth of the dip decreases with the increasing of $n_{en}$, which can be explained that more lights will be leaked to the outside of the fiber when $n_{en}$ increases. When the RI of the solution is larger than 1.3394, the transmission depth becomes less than 4.5 dB, which is not large enough for wavelength shift measurement. Therefore, the measurable RI range of the device is from 1.3324 to 1.3394. The relationship between the resonant dip wavelength $\lambda_{dip}$ and the RI $n_{en}$ of the solution is plotted in Figure 5. It can be seen that $\lambda_{dip}$ decreases linearly from 1551.2 to 1544.8 nm when $n_{en}$ increases from 1.3324 to 1.3394, so the experimental sensitivity $S$ of the device is 914 nm/RIU. The resonant dip wavelength $\lambda_{dip}$ is observed to be at 1522 and 1560 nm for the bent structure with $R = 420$ and 500μm, the corresponding sensitivity $S$ are 603 and 1031nm/RIU, and the measurable RI range are 1.3324-1.3407, 1.3324-1.3435, respectively. For the bent structure with $R = 1000\mu m$, the sensitivity obtained in the experiment is almost zero. The reason can be explained that, for a bent structure with larger bending radius, the cladding mode is harder to be effected by the surrounding environment. Figure 6 plot the relationship between the sensitivity and the bent length, and the relationship between the resonant dip and the bent length. It can be seen that the sensitivity of the sensor and the resonant dip both linearly increase with the increasing of the bending radius, which is consistent with the theoretical analyze.

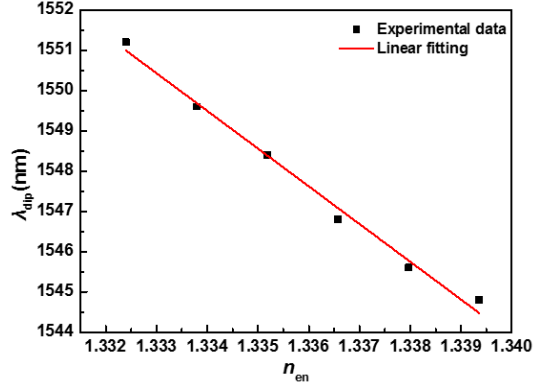

FIG. 5. The relationship between the resonant dip wavelength $\lambda_{dip}$ and the RI $n_{en}$ of the solution. Points represent the experiment results and red line is the linear fitting.

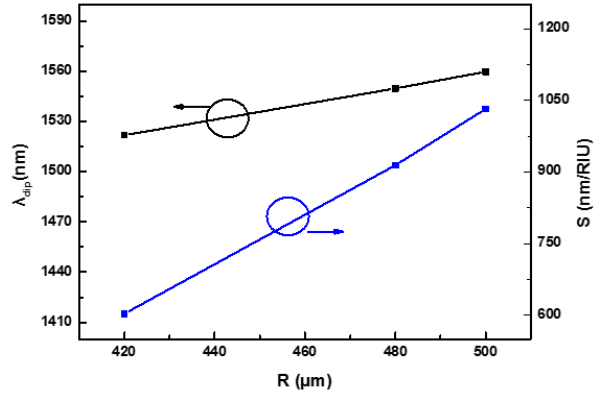

FIG. 6. The sensitivity and the resonant dip versus the bending radius.

In conclusion, a fiber sensor based on a semicircular bent fiber has been demonstrated. In the bending region, part of the core mode is coupled to the cladding mode. After leaving the region, the cladding mode will be coupled back to the core mode. Because the ERI of the core mode and the cladding mode are different, interference will be formed at the output end of the fiber. Both the theoretical and experimental results show that the resonant dip wavelength and the sensitivity of the bent structure to the RI of the surrounding environment increase linearly with the increasing of the bending radius of the structure, and the resonant dip wavelength shows a blue shift as the increasing of the RI of the surrounding environment. According to the experiment, a sensitivity as high as 1031 nm/RIU was obtained for the bent structure with bending radius of 500 μm for the RI of the surrounding environment in a range of 1.3324 to 1.3435. Due to its easy fabrication method and high sensitivity, we anticipate it will have potential applications for bio-sensing and chemical sensing.